\documentclass{websci2012}
\usepackage{times}
\usepackage{amsmath}

\begin{document}

\setlength{\paperheight}{11in}
\setlength{\paperwidth}{8.5in}
\setlength{\pdfpageheight}{\paperheight}
\setlength{\pdfpagewidth}{\paperwidth}


\title{Semantic Networks of Interests in Online NSSI Communities}
\numberofauthors{2}
\author{
  \alignauthor Dmitry Zinoviev, Dan Stefanescu\\
    \affaddr{Mathematics and Computer Science Department}\\
    \affaddr{Suffolk University, Boston MA, USA}\\
    \email{\{dzinoviev,dstefanescu\}@suffolk.edu}
  \alignauthor Lance Swenson, Gary Fireman\\
    \affaddr{Psychology Department}\\
    \affaddr{Suffolk University, Boston MA, USA}\\
    \email{\{lswenson,gfireman\}@suffolk.edu}
}

\maketitle

\begin{abstract}
Persons who engage in non-suicidal self-injury (NSSI), often conceal
their practices which limits the examination and understanding of
those who engage in NSSI. The goal of this research is to utilize
public online social networks (namely, in LiveJournal, a major
blogging network) to observe the NSSI population’s communication in a
naturally occurring setting. Specifically, LiveJournal users can
publicly declare their interests. We collected the self-declared
interests of 22,000 users who are members of or participate in 43
NSSI-related communities. We extracted a bimodal socio-semantic
network of users and interests based on their similarity. The semantic
subnetwork of interests contains NSSI terms (such as ``self-injury''
and ``razors''), references to music performers (such as ``Nine Inch
Nails''), and general daily life and creativity related terms (such as
``poetry'' and ``boys''). Assuming users are genuine in their
declarations, the words reveal distinct patterns of interest and may
signal keys to NSSI.
\end{abstract}




\section{Introduction}

Non-suicidal self-injury (NSSI) is the direct, deliberate destruction
of one's own body tissue in the absence of suicidal
intent~\cite{nock2009}. It is practiced primarily by adolescents and
young adults~\cite{jacobson2007} 
and is often concealed from others. 
Common NSSI activities include skin cutting, banging or hitting oneself, and
burns~\cite{klonsky2008}.

Recent prevalence estimates suggest that 14\% to 21\% of adolescents
and 17\% to 25\% of young adults 
have engaged in NSSI at some point in their
lives~\cite{klonsky2008, ross2002}. Furthermore, NSSI is repeatedly
found to be associated with significant emotional and behavioral
dysfunction (e.g., eating disorders, suicide
~\cite{nock2009-1}). These findings highlight the need to enhance
understanding and prevention of NSSI and its psychiatric sequalea.

The goal of this research is to find mechanisms that could identify
NSSI persons by automatically analyzing secondary data publicly
available from massive online social networks (MOSN), without
explicitly interacting with the subjects. Many popular MOSNs (e.g.,
Facebook and LiveJournal) allow users to declare their interests,
either explicitly or in the form of ``likes.'' While these interests
are often selected randomly and polluted with ``status words,'' we
found a very significant correlation between interest lists and
membership in NSSI online communities in at least one major
MOSN---LiveJournal~\cite{Livejournal,Zakharov07}, a blogging social
network.

This association between interest lists and NSSI community membership
suggests that ``likes'' or interest lists may be serving as identity
signals ``communicating aspects of individuals (e.g., group membership
or other preferences) to others in the social
world''~\cite{berger2008}. Such identity signals gain greater meaning
(i.e., signal value) as their association with group membership
strengthens. From an identity-signaling perspective, identity signals
with greater signal value can influence others, particularly others
who aspire for group membership, to adopt behaviors characteristic of
the larger group~\cite{berger2008}.

To investigate the value of interest lists generated by members of
NSSI online communities in LiveJournal, we used the declared interests
as nodes and similarities between their users as edges to build a
semantic network. The layout of the network consists of four clearly
separated word clusters, one of which corresponds to the pathological
terms (e.g. ``self-injury'' and ``razor'') and the other three refer
to daily life, popular music, and creativity. We expect that the
bridge terms that connect the pathology cluster with the remaining
three clusters can be used as beacons signaling the potential presence
of an NSSI behavior.

The rest of the paper is organized as follows: in Section~1, we
describe the data acquisition process; Sections~2 and 3 explain the
semantic network generation and the resulting network organization;
network comparative assessment is presented in Section~4; in
Section~5, we conclude and outline our future research plans.

\section{Data Collection}\label{data}
Our analysis is based on the data set collected from Live\-Journal---a
popular massive online blogging social network site (BSN). A BSN
allows individual bloggers to form contact lists, subscribe to their
friends' blogs, comment on selected blog posts, declare interests, and
participate in com\-munities---collective blogs. Thus, a blogging
network is a bimodal venue where users engage in both publishing and
social activities~\cite{zinoviev2012}. As of Spring 2012, LiveJournal
has 32 mln individual and community accounts. A LiveJournal user
maintains his/her personal blog (public or private) and may be a
member of an unlimited number of special- and general-interest
communities.
 
We identified 43 NSSI-related communities in LiveJournal\footnote{A
  complete list of communities with their posting and membership
  statistics, etc. is available from the authors in electronic form by
  email.}. Users are associated with the communities either explicitly
(by membership) or implicitly (by posting to the community blogs
without becoming formal members, where permitted). Some of these
communities promote NSSI activities, while others advocate for NSSI
abstinence.

We collected all self-declared interests of the 22,000 LiveJournal
users associated with the selected communities (by membership or by
posting, as described above). The total number of harvested interests
is $\sim$150,000, including misspelled, abbreviated, and hyphenated
variants.

Thus, we formed a matrix $M$ where $M_{ij}=1$ iff the user $U_i$ has
declared the interest $V_j$, and $M_{ij}=0$, otherwise. In other
words, $M$ is the adjacency matrix of a two-mode network of users and
their interests.
 
\section{Semantic Network Generation}\label{generation}
We use the matrix $M$ to generate a semantic
network~\cite{shapiro1992} of interests corresponding to the NSSI
population. This network is a one-mode projection of the original
two-mode network induced by the matrix $M$. It is undirected,
weighted, and signed. The nodes of the network represent interests
$I_i$ and the edges represent the corresponding general similarities
$C_{ij}\in[-1,1]$.

Thematic (e.g., NSSI-related) communities are more homogeneous than
general-interest communities. They consist of people who are similar
in a certain sense. In an extreme case, all community members would be
uniformly interested in the community subject and use common
terminology. This similarity should be taken into consideration while
calculating correlations between declared interests. It has been shown
by Kovacs~\cite{kovacs2010} and confirmed by our finding that agent
agnostic Pearson correlation underestimates the proximity of
terms. Kovacs generalized similarity measures take the population
structure into account. They are defined recursively: two terms are
similar with correlation $\Theta_{ij}$ if they are used by similar
people; two people are similar with correlation $\Phi_{ij}$ if they
use similar terms ($\Theta_{ij},\Phi_{ij}\in[-1,1]$). 

Let $M_i=M_{i,\cdot}-\overline{M_{i,\cdot}}$ and
$M^j=M_{\cdot,j}-\overline{M_{\cdot,j}}$ be the $i$'th row or the
$j$'th column of the matrix $M$, respectfully, centered by subtracting
the mean of the corresponding row or column. Then matrices $\Theta$
and $\Phi$ can be calculated recursively by starting with two
appropriately sized identity matrices $I$:
\begin{align*}
  \Theta_0&=I, \Phi_0=I,\\
  \Theta_{ij,k+1}&=
  M_i\Phi_k M^T_j/\sqrt{\left(M_i \Phi_k M^T_i\right)\left(M^j\Phi_k
    M^{jT}\right)},\\
  \Phi_{ij,k+1}&=M^T_i \Theta_k M_j/\sqrt{\left(M^T_i \Theta_k
    M_i\right)\left(M^{jT} \Theta_k M^j\right)}.
\end{align*}
\noindent
After a number of iterations the algorithm converges to the ``true''
values of $\Phi\approx\Phi_\infty$ and $\Theta\approx
\Theta_\infty$. The similarities between community members $\Phi$,
though calculated, are not used in this study. 

As a side note, in the case of totally heterogeneous population,
$\Theta=\Sigma$ and $\Phi=\text{I}$ (each person is similar only to
herself).

By construction, $\Theta$ is a dense symmetric signed square matrix
with few or no zero terms. The distribution of similarity measures in
the matrix is close to uniform. The similarities in the matrix are
sustained by the whole body of interests and are robust against random
variations of individually declared interests.

Calculating $\Theta$ for 150,000 interests is computationally
infeasible due to time constraints and arithmetic imprecision. We
restricted our study to the top 600 most often declared interests
shared by $\sim$14,000 NSSI bloggers. That was the largest matrix that
could be evaluated on a 64-bit AMD desktop computer with 8GB of RAM.

\section{Semantic Network Analysis}\label{organization}
To explore the organization of the semantic network of interests, we
extracted some of the strongest generalized similarities between the
interests by creating another adjacency matrix $\Psi$:
\begin{equation}
  \Psi_{ij}=
  \begin{cases}
    \Theta_{ij}&\text{if $\Theta_{ij}\ge0.8$}\\
    0&\text{else}
  \end{cases}
  \label{psimatrix}
\end{equation}
Matrix $\Psi$ is square, sparse (its density is 12\%), symmetric,
undirected, weighted (in a limited range), and unsigned. It has 42,000
non-zero entries that correspond to 21,000 network edges.

\begin{figure}[tb!]\centering
\strut\epsfig{file=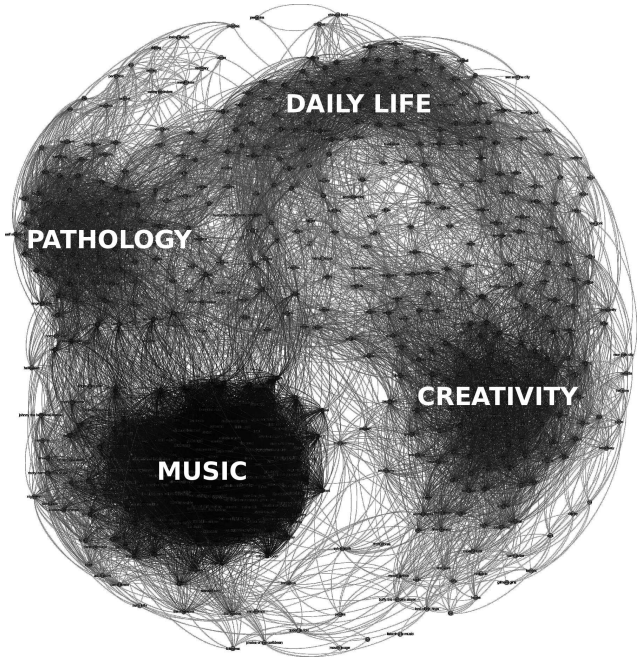,width=\columnwidth}
\caption{\label{map}Semantic network of interests in the NSSI-related
  communities}
\end{figure}

We used program Gephi~\cite{bastian2009} to visualize the network
described by matrix $\Psi$. The sketch of the network is shown in
Figure~\ref{map}\footnote{The detailed network map is available from
  the authors in electronic form by email.}. The network has a clear
hierarchical structure. It consists of four major clusters: ``music''
(MUS), ``pathology'' (PAT), ``daily life/emotions'' (DLE), and
``creativity'' (CRE). Some most frequently declared interests from
each of the clusters are shown below:

\begin{description}
\item[MUS:]atreyu, him, incubus, korn, my chemical romance, nirvana,
  rancid, system of a down, the perfect circle;
\item[PAT:]alcohol, anorexia, bulimia, burning, cutting, handcuffs,
  pain, self-injury and self-mutilation (both with and without the
  dash), spikes, weeds;
\item[DLE:]cameras, cloths, dvds, flirting, flowers, fun, quotes,
  smiling, hearts (also as an HTML entity \&hearts; and as
  $\heartsuit$);
\item[CRE:]astrology, books, languages, philosophy, psychology,
  shakespeare, sociology, travel, wine.
\end{description}

There is surprisingly little connectivity between the clusters CRE and
MUS. The remaining border zones are spanned with few important bridge
interests:
\begin{description}
\item[PAT/MUS:] (black) eyeliner, girl interrupted, metal;
\item[PAT/DLE:] candy, girls, insomnia, red, rock music, sex;
\item[MUS/DLE:] animals, camping, fashion, games, honesty, humor,
  travel(l)ing;
\item[All four clusters:] bands, bracelets, hoodies, lesbians, making
  out.
\end{description}
Since, from the point of view of the NSSI users, the bridge terms are
similar both to pathological and non-pathological terms, their
occurrence in a text may be a signal of a potentially NSSI behavior.

The presence of the tightly interwoven MUS component is equally
surprising, given that music is not an explicit topic in any of the
NSSI communities.

\section{Semantic Network Comparative Assessment}\label{calibration}
While many of the associations shown in the map in Figure~\ref{map}
may be specific to the NSSI community members, some may be either
totally random or specific to the age or cultural group to which these
members belong. Thus, some associations that seemingly suggest NSSI
behavior, may turn out to be misleading.

To identify NSSI-specific information, we attempted to compare the NSSI
communities to a random sample of LiveJournal users whose declared age distribution was similar  to that of NSSI users. 
The members of the chosen sample seem to share very few interests with the NSSI population under study.
This is not
surprising, given that the $\sim$32 mln LiveJournal users belong to
different ethnic and cultural groups and, when chosen at random, are
unlikely to share interests expressed by a small number of words form a very limited vocabulary.

Next we tried to find appropriate  communities that may share interests with the NSSI population even given a very limited power of expression.
After some research we identified two LiveJournal communities that
have demographics similar to the NSSI communities and focus on
non-pathological topics: ``sexy-mood-music'' (SMM, 6,700 members,
average age 25 years) and ``movies-in-fifteen-minutes'' (M15M,
13,300 members, average age 28 years). These communities cater
to music and video fans, respectively. 

We collected the top 450 and 550 most frequently used interests of the
SMM and M15M members and used the technique described above to
generate their semantic maps $\Psi_\text{SMM}$ and
$\Psi_\text{M15M}$. We then calculated the intersection between the
NSSI semantic network and each of the other semantic networks under
consideration. The intersection contains the associations that are
significant for both communities and presumably are pathology-free.

We combine semantic network edges using fuzzy set theoretical
operations for intersection and difference:
\begin{align}
A\cap B&=\min\left(\alpha,\beta\right)\label{inter}\\
A\,\backslash B&=\min\left(\alpha,1-\beta\right)\label{fisher}
\end{align}
Here, $A$ and $B$ are edges, and $\alpha$ and $\beta$ are generalized
similarities associated with the edges. 

Let $\Psi\cap x$ be the intersection of the original network and a
comparison network $x\in\{\text{SMM},\text{M15M}\}$. Then for each
edge $e_{ij}=(V_i, V_j)$ in $\Psi$, if the corresponding edge also
exists in $x$, then this edge is added to the intersection network
with the weight calculated using Eq.~\ref{inter}. Otherwise, its
weight is taken to be 0 to emphasize the lack of either similarity
between the two terms:
\begin{equation}
  \left(\Psi\cap x\right)_{ij}=
  \begin{cases}
    \Psi_{ij}\cap x_{ij}&\text{if $e_{ij}\in \left(\Psi \cap
      \text{x}\right)$}\\ 0&\text{else}
  \end{cases}
  \label{sharedmatrix}
\end{equation}
In other words, if two terms are considered substantially similar both
in $\Psi$ and in $x$, then their similarity is universal with respect
to $\Psi$ and $x$, that is, neither $\Psi$- nor $x$-specific.

\begin{figure}[tb!]\centering
\strut\epsfig{file=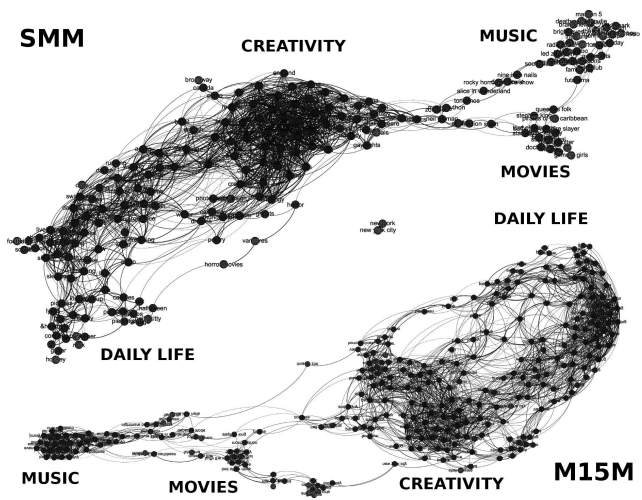,width=\columnwidth}
\caption{\label{map-common}Two intersection networks showing common
  interests $\text{NSSI}\cap\text{SMM}$ and
  $\text{NSSI}\cap\text{M15M}$}
\end{figure}

The resulting intersection networks for
$x\in\{\text{SMM},\text{M15M}\}$ have fewer nodes (300 and 220) and
about the same density (9\% and 12\%). We analyzed them using Gephi
software (Figure~\ref{map-common}) and discovered that they have a
remarkably common structure: both networks have dense and strongly
connected CRE and DLE clusters and smaller and less connected to the
``mainland'' music/movies clusters. The latter clusters, in turn,
consist of easily identifiable ``movies'' (MOV) and MUS subcomponents.

The intersection networks with their structural similarity exhibit many characteristics observed in the
typical adolescent
development~\cite{larson2002}.

Next we adjust the original network with respect to the comparison
networks as a way to better distinguish the features of the NSSI
semantic network.

Let $\Psi\backslash x$ be the difference between the original network
$\Psi$ and a comparison network $x$. If edge $e_{ij}$ exists in $\Psi$
but not in $x$, then it is essential and is inserted in the difference
network with its original weight. If the edge exists in both networks,
it is inserted with the weight calculated using
Eq.~\ref{fisher}. Otherwise, the edge exists only in $x$; it is
irrelevant and is not inserted:
\begin{equation}
  \left(\Psi\,\backslash x\right)_{ij}=
  \begin{cases}
    \Psi_{ij}\,\backslash x_{ij}&\text{if $e_{ij}\in \Psi$}\\
    0&\text{else}
  \end{cases}
  \label{compelementmatrix}
\end{equation}

\begin{figure}[tb!]\centering
\strut\epsfig{file=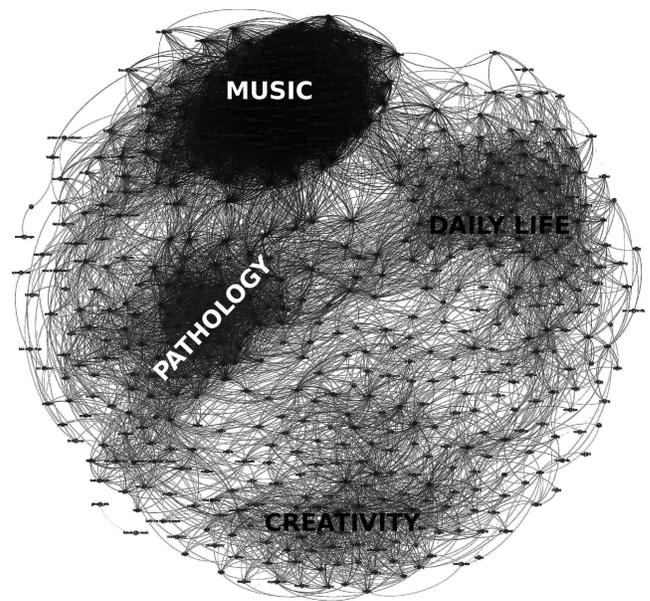,width=\columnwidth}
\caption{\label{map-calibrated}Semantic network of interests in the
  NSSI-related communities, adjusted for common interests;
  cf.~Figure~\ref{map}}
\end{figure}

In other words, if two terms are considered substantially similar in
$\Psi$ but not in $x$, then their similarity is $\Psi$-specific but
not $x$-specific. These terms may be perceived as similar by the NSSI
users because of their NSSI pathologies.

The adjusted NSSI interest network with respect to $x=\text{M15M}$ is
shown in Figure~\ref{map-calibrated}. The new network has much less
dense CRE and DLE clusters; cf. the original network in
Figure~\ref{map}. The PAT and especially MUS clusters are still very
dense. From Figure~\ref{map-calibrated}, we can identify two groups of
possible NSSI beacon interests:

\begin{description}
\item[Non-PAT interests in PAT:] angelina jolie, bdsm, beer, being
  alone, bisexuality, black, boots, crying, dying, fire, fishnets,
  goth(ic), graveyards, hair dye, horror, industrial, lust,
  perfection, porn, serial killers, sex, tattoos, tears, vampires,
  wicca, witchcraft, etc.
\item[Bridge interests:] anxiety, bracelets, corsets, edgar allen poe,
  emotions, (black) eyeliner, girl interrupted, girls, glitter, horror
  movies, insomnia, leather, lesbians, magick ({\it sic}), marylin
  monroe, (heavy) metal, night, poems, red, safety pins, screaming,
  spikes, techo, tori amos, etc.
\end{description}

Our findings also appear indicative of the growing global middle-class
youth culture revolving around leisure activities (e.g., music, art)
reflecting adolescent development in internationally-connected
networks~\cite{larson2002}. This is further supported by the
similarities between the NSSI interested communities and the
non-pathological comparison communities. Notably both sets of
communities included entertainment, creativity, and daily life
clusters.

\section{Conclusion and Future Work}\label{conclusion}
Exposure to NSSI via Internet use (e.g., MOSNS, YouTube) may
facilitate the adoption and maintenance of NSSI among vulnerable
individuals via social contagion processes~\cite{lewis2011,whitlock2009}. With this in mind, we constructed a semantic
network of interests declared by non-suicidal self-injury (NSSI)
bloggers of LiveJournal. The network consists of four clearly
separated interest clusters corresponding to the pathological terms
(e.g. ``self-injury'' and ``razor''), daily life, popular music, and
creativity. The interests that bridge gaps between the pathology
cluster and the other three clusters can be used as beacons signaling
the potential presence of an NSSI behavior. These bridge interests
appear to be valuable identity signals~\cite{berger2008} serving as
linkages between NSSI group membership and larger youth
culture. Future research targeting individuals’ use of these bridge
terms as a means to identify NSSI-oriented social groups would further
support this interpretation and could inform prevention efforts aimed
at early identification of vulnerable individuals at risk for NSSI.

In related research individuals with a history of NSSI are found to
view themselves negatively (e.g., less intelligent and more
emotionally unstable) and as having lower social capital (e.g., less
attractive, weak social skills~\cite{claes2010}). The extent of MOSN
NSSI-related communities on LiveJournal could evidence the limited
opportunities for social networking among people (e.g., self-harmers)
who find themselves excluded from their local communities/local peer
networks. Future work is needed examine how members of NSSI-related
communities use MOSNs to affirm a sense of meaning and obtain social
support and expanding social capital. At the same time, increased time
in unstructured peer interactions via NSSI-related MOSNs may lead to
further involvement in deviant and antisocial behavior in early
adulthood~\cite{larson2002}.

In this study, we considered only self-declared interests displayed on
the bloggers' profile pages. Some of these interests may have been
chosen randomly or based on certain (sub)cultural considerations, and
do not necessarily reflect the real user's attractions. As the next
step in this direction, we plan to study keywords in the messages
posted to the NSSI communities. We expect that the free-form language
of the messages is a better proxy for the pathological behavior. If
our hypothesis is right, then the semantic network generated from the
keywords will differ from the network $\left(\Psi\backslash x\right)$
constructed in this study. The area of overlap will probably be the
cluster of ``true'' NSSI interests.

\section*{Acknowledgment}
This research has been supported in part by the College of Arts and
Sciences, Suffolk University, through an undergraduate research
assistantship grant. The authors are grateful to Zo\"e Wells of
Suffolk University for preliminary data collection and Dr.~Jim
Hollander and Prof.~John Boyd for suggestions on combining graphs.

\bibliographystyle{acm-sigchi}
\bibliography{cs}

\end{document}